\documentclass[prl,aps,twocolumn,showpacs,floats]{revtex4}
\usepackage{graphicx,amsmath,epsf}

\renewcommand{\vr}{\mathbf{r}}

\begin{document}

\title{Interaction correction to the conductance of a ballistic conductor}

\author{Piet W.\ Brouwer and Joern N.\ Kupferschmidt}

\affiliation{Laboratory of Atomic and Solid State Physics, Cornell
  University, Ithaca, NY 14853, USA}
\affiliation{Arnold Sommerfeld Center for Theoretical Physics, Ludwig-Maximilians-Universit\"at, 80333 M\"unchen, Germany}

\date{\today}


\begin{abstract}
In disordered metals, electron-electron interactions are the origin of
a small correction to the conductivity, the ``Altshuler-Aronov
correction''. Here we investigate the Altshuler-Aronov
correction $\delta G_{\rm AA}$ of a conductor in which the electron
motion is ballistic and chaotic. We consider the case of a double
quantum dot, which is the simplest example of a ballistic
conductor in which $\delta G_{\rm AA}$ is nonzero. The fact that the
electron motion is ballistic leads to an exponential suppression of 
$\delta G_{\rm AA}$ if the
Ehrenfest time is larger than the mean dwell time $\tau_{\rm D}$ or
the inverse temperature $\hbar/T$. 
\end{abstract}

\maketitle

There are two quantum corrections of comparable magnitude to the
conductivity of a disordered normal metal at low temperatures: the
weak localization correction, which has its origin in the 
constructive interference of electrons traveling along time-reversed 
paths \cite{kn:anderson1979,kn:gorkov1979},
and the ``Altshuler-Aronov correction'', which is caused by
electron-electron 
interactions \cite{kn:altshuler1979a}.
These two corrections can be
distinguished through their different dependences on 
temperature and magnetic field.
For mesoscopic conductors, which are characterized by their
conductance, not their conductivity, weak localization and the
Altshuler-Aronov correction represent small changes to the mean
conductance after taking an average over different disorder 
configurations \cite{kn:altshuler1995}.

Although weak localization was discovered in the context of disordered
metals, it also occurs if the electron motion is
ballistic and the only source of scattering is specular reflection off
sample boundaries or artificial macroscopic scattering
sites \cite{kn:marcus1992,kn:baranger1993,foot1}. 
There is an important difference between the two cases,
however: For weak localization in ballistic conductors
a key role is played by the Ehrenfest time 
$\tau_{\rm E}$ \cite{kn:aleiner1996}, a time which has 
no counterpart in disordered metals. The Ehrenfest time
separates regimes of classical-deterministic and 
quantum-probabilistic motion \cite{kn:larkin1968,kn:zaslavsky1981},
thus serving
as a short-time ``threshold'' for quantum corrections in a ballistic 
conductor. The weak localization correction
$\delta G_{\rm WL}$ has an exponential dependence on $\tau_{\rm E}$ if
$\tau_{\rm E}$ is
large, $\delta G_{\rm WL} \propto \exp(-\tau_{\rm E}/\tau_{\rm D} -
\tau_{\rm E}/\tau_{\phi})$, 
where $\tau_{\rm D}$ is the dwell time and $\tau_{\phi}$ the dephasing
time \cite{kn:aleiner1996,kn:adagideli2003,kn:altland2007}.

In this letter we show that $\tau_{\rm E}$ also serves as a
short-time threshold for the Altshuler-Aronov correction
$\delta G_{\rm AA}$ in a ballistic conductor, in a manner quite
similar to the way it appears in the theory of weak
localization and other quantum corrections that do not rely on
electron-electron interactions. In particular, we show that $\delta
G_{\rm AA}$ has an exponential dependence on $\tau_{\rm E}$ if
$\tau_{\rm E}$ is large,
\begin{equation}
  \delta G_{\rm AA} \propto e^{-\tau_{\rm E}/\tau_{\rm D}-2
\pi T \tau_{\rm E}/\hbar},
  \label{eq:exponential}
\end{equation}
where 
$T$ is the
temperature. The exponential sensitivity to temperature is special to
the interaction correction; Although weak localization 
depends on $T$ implicitly via the temperature-dependence of
the dephasing time, the corresponding exponential dependence is much
weaker since $\tau_{\phi} \gg \hbar/ T$ \cite{kn:altshuler1995}.

For most studies of quantum corrections in ballistic conductors, the
geometry of choice is a ballistic cavity or ``quantum dot''.
This geometry is not suitable for a theory
of the Altshuler-Aronov correction, however, because
$\delta G_{\rm AA} = 0$ in a ballistic quantum
dot \cite{kn:brouwer1999c,kn:brouwer2005}. Therefore,
we here calculate $\delta G_{\rm AA}$ for a ``double quantum dot'', a
device consisting of two quantum dots coupled by a ballistic
contact. Although $\delta G_{\rm AA}$ will be quantitatively different in other geometries, we believe that the qualitative features mentioned above carry over to the general case.

\begin{figure}
\epsfxsize=0.95\hsize
\epsffile{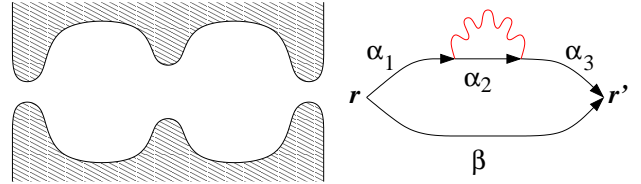}
\caption{\label{fig:1} (Color online)
  Schematic drawing of a double quantum
  dot (left), with a generic set of four trajectories that
  contributes to the interaction correction to the
  conductance (right). The wiggly line represents the interaction
  propagator.
}
\end{figure}

A schematic drawing of a double quantum dot is shown in
Fig. \ref{fig:1}. The two quantum dots are connected to source and
drain reservoirs via ballistic contacts of conductances $G_1$, $G_2
\gg e^2/h$ respectively; They are connected with each other via a
ballistic contact with conductance $G_{\rm c}$. The classical electron 
dynamics in each quantum dot is ballistic and chaotic, with Lyapunov
exponent $\lambda$. (We assume the same Lyapunov exponent for both
quantum dots.) The Ehrenfest time then reads
\begin{equation}
  \tau_{\rm E} = \lambda^{-1} \ln(k_F L),
  \label{eq:tauEdef}
\end{equation}
where $k_F$ is the Fermi wavenumber and $L$ the dot size. We take the
semiclassical limit $k_F L \gg 1$, so that the logarithm in Eq.\
(\ref{eq:tauEdef}) is large. For a ballistic double quantum dot, the
disorder average is replaced by an average over the Fermi energy or
over variations of the dot's shape.

In the semiclassical limit the 
electron-electron interaction in the double quantum dot takes a
particularly simple form: It is determined by 
the capacitances $C_{1,2}$ of the two dots and their mutual 
capacitance $C_{\rm c}$ only. (For the case of a single quantum dot the
corresponding interaction is known as the ``universal interaction
Hamiltonian'' \cite{kn:aleiner2002}.) As a result, the interaction
propagator ${\cal D}(\vr_1,\vr_2;\omega)$ for the double dot
is spatially homogeneous inside each dot, so that it 
may be represented by a $2 \times 2$ 
matrix $\tilde{{\cal D}}$, where the matrix indices refer to the 
two quantum dots,
\begin{eqnarray}
  \tilde{{\cal D}}^{\rm R}(\omega)^{-1} &=&
  - {\tilde C}/e^2 -
  [\tilde{\nu}^{-1} - i e^2 \omega (\hbar \tilde {G})^{-1}]^{-1}
,
  \label{eq:Dprop}
\end{eqnarray}
with $\tilde{C}_{mn} = C_m \delta_{mn} + C_{\rm c}(-1)^{m+n}$,
$\tilde G_{mn} = G_m \delta_{mn} + G_{\rm c} (-1)^{m+n}$, and
$\tilde \nu_{mn} = \nu_m \delta_{mn}$, where $\nu_1$ and $\nu_2$ are
the level densities in each dot. For all frequencies of interest one
may neglect the first term in Eq.\ (\ref{eq:Dprop}) and approximate
$\tilde {\cal D}^{\rm R}(\omega) = i e^2 \omega (\hbar \tilde G)^{-1}
- \tilde \nu^{-1}$.


The Altshuler-Aronov correction $\delta G_{\rm AA}$ can be calculated
from the interaction correction to the single-electron Green function
${\cal G}(\vr,\vr';\omega)$ (before the ensemble average). 
Without interactions, the relation
between ${\cal G}$ and 
the conductance $G$
is given by the Kubo formula,
\begin{equation}
  G = \frac{e^2 \hbar}{\pi} \hat v_x \hat v_{x'}
  \int dy dy' 
  \int d\xi\,
  \frac{{\cal G}^{\rm R}(\vr,\vr';\xi) {\cal G}^{\rm
  A}(\vr',\vr;\xi)}{4 T \cosh^2(\xi/2 T)},
  \label{eq:Kubo}
\end{equation}
where $x$ and $x'$ ($y$ and $y'$) are longitudinal (transverse) 
coordinates in the source and drain contacts, respectively, 
and 
\begin{eqnarray*}
  \hat v_x {\cal G}(\vr,\cdot) {\cal
  G}(\cdot,\vr) &=& \frac{\hbar e}{2 m i} [(\partial_x
  {\cal G}(\vr,\cdot)) {\cal G}(\cdot,\vr) 
  \nonumber \\ && \ \ \mbox{}
  - {\cal G}(\vr,\cdot) \partial_x {\cal
    G}(\cdot,\vr)].
\end{eqnarray*}
The Altshuler-Aronov correction $\delta G_{\rm AA}$ then follows
from Eq.\ (\ref{eq:Kubo}) if one makes the substitution ${\cal G}^{\rm
  R,A} \to {\cal G}^{\rm R,A} + \delta {\cal G}^{\rm R,A}$, 
with 
\cite{kn:altshuler1979a,kn:aleiner1999}
\begin{widetext}
\begin{eqnarray}
  \delta {\cal G}^{\rm
  R}(\vr,\vr';\xi) &=&
  \int \frac{d\omega}{4 \pi i} \tanh\left( \frac{\omega - \xi}{2 T}
  \right)
  \int d\vr_1 d\vr_2 {\cal G}^{\rm R}(\vr,\vr_1;\xi) 
  {\cal G}^{\rm R}(\vr_2,\vr';\xi) 
  \nonumber \\ && \mbox{} \times
  \left\{
  {\cal D}^{\rm R}(\vr_1,\vr_2;\omega)
  {\cal G}^{\rm A}(\vr_1,\vr_2;\xi -
  \omega) 
  -
  {\cal D}^{\rm A}(\vr_1,\vr_2;\omega)
  {\cal G}^{\rm R}(\vr_1,\vr_2;\xi -
  \omega) 
  \right\},
  \label{eq:deltaGreen}
\end{eqnarray}
\end{widetext}
and a similar expression for $\delta {\cal G}^{\rm A}$.
Notice that Eq.\ (\ref{eq:deltaGreen}) represents the Fock 
contribution to 
$\delta {\cal G}$ only. The Hartree contribution to $\delta {\cal G}$ 
vanishes because of the special form of the interaction 
${\cal D}(\vr_1,\vr_2;\omega)$, see Eq.\ (\ref{eq:Dprop}) above.

Together, Eqs. (\ref{eq:Dprop})--(\ref{eq:deltaGreen}) express
$\delta G_{\rm AA}$ as an 
integral over a product of four single electron Green functions. 
What remains to be done is to calculate the ensemble average. 
Hereto, we use a semiclassical approach inspired by the
semiclassical calculation of the weak localization correction to the
conductance in Ref.\ \cite{kn:richter2002}. First, we use the
standard expression of the Green
function ${\cal G}(\vr,\vr';\xi)$ as a sum over classical
trajectories $\alpha$ connecting the points $\vr$ and
$\vr'$ \cite{kn:gutzwiller1990},
\begin{eqnarray}
  {\cal G}^{\rm R}(\vr,\vr';\xi) &=&
  {\cal G}^{\rm A}(\vr',\vr;\xi)^* \nonumber \\ &=&
  \frac{2 \pi}{(2 \pi i \hbar)^{3/2}}
  \sum_{\alpha} A_{\alpha} e^{i {\cal S}_{\alpha}(\xi)/\hbar},
  \label{eq:GreenSemi}
\end{eqnarray}
where $A_{\alpha}$ is the stability amplitude of $\alpha$ and ${\cal
S}_{\alpha}$ its classical action. Actions at different
energies are related via 
\begin{equation}
  {\cal S}_{\alpha}(\xi) -
  {\cal S}_{\alpha}(\xi - \omega) = \omega \tau_{\alpha},
\end{equation}
where $\tau_{\alpha}$ is travel time from $\vr'$ to
$\vr$ along $\alpha$.
\begin{figure}
\epsfxsize=0.9\hsize
\hspace{0.05\hsize}
\epsffile{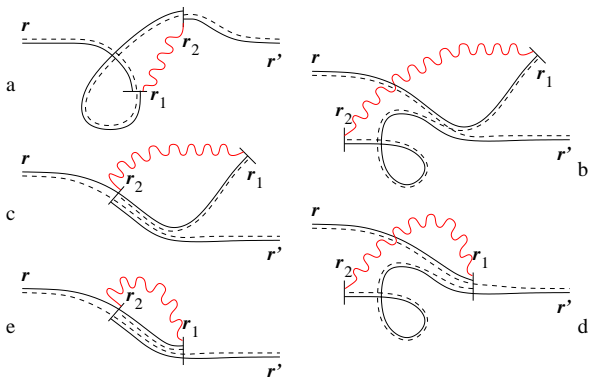}
\caption{\label{fig:2} (Color online)
Schematic drawing of five configurations of
classical trajectories contributing to $\delta G_{\rm AA}$. 
Solid (dashed) trajectories
represent retarded (advanced) Green functions. 
Five more trajectory
configurations are obtained by interchanging the roles of retarded and
advanced Green functions. The wiggly lines represent the advanced
interaction propagator.}
\end{figure}
Substitution of Eq.\ (\ref{eq:GreenSemi}) into Eqs.\
(\ref{eq:Kubo}) and (\ref{eq:deltaGreen}) expresses $\delta G_{\rm
  AA}$  as a sum
over four classical trajectories $\alpha_1$, $\alpha_2$,
$\alpha_3$, and $\beta$ [Fig.\ \ref{fig:1}, right panel]. 
For each set of four classical trajectories there are four
contributions to $\delta G_{\rm AA}$, corresponding to
the two terms in Eq.\ (\ref{eq:deltaGreen})
and the two terms in the corresponding expression for $\delta
{\cal G}^{\rm A}$.
The four cases differ by the assignment which of the trajectories 
$\alpha_1$, $\alpha_2$,
$\alpha_3$, and $\beta$ correspond to retarded and which correspond to
advanced Green functions.
Each configuration of trajectories contributes to the average
conductance $\langle G \rangle$ only if the total action difference
$\Delta {\cal S}$
is of order $\hbar$ systematically. This occurs only if the
`retarded' and `advanced' trajectories are piecewise equal, up
to classical phase space distances of order
$\hbar^{1/2}$ and below \cite{kn:aleiner1996,kn:richter2002}.
Five possible configurations of classical
trajectories that meet this requirement are shown in Fig.\
\ref{fig:2}. In Fig.\ \ref{fig:2}a, $\alpha_1$, $\alpha_2$, and
$\alpha_3$ represent retarded Green functions, whereas $\beta$
represents an advanced Green function. 
In Fig.\ \ref{fig:2}b--e, $\alpha_1$ and
$\alpha_3$ represent retarded Green functions, whereas $\alpha_2$ and
$\beta$ represent advanced Green functions. There are five more
trajectory configurations that contribute to $\delta G_{\rm AA}$ which
can be obtained from those of Fig.\ \ref{fig:2} by interchanging the
roles of advanced and retarded Green functions.

In Fig.\ \ref{fig:2}a, the points $\vr_1$ and $\vr_2$ are on the
trajectory $\beta$ (up to a quantum uncertainty), 
and the three trajectories $\alpha_1$, $\alpha_2$,
and $\alpha_3$ are equal to different successive segments of $\beta$
(again up to quantum uncertainties). In Figs.\ \ref{fig:2}b--e, the situation
is more complicated because the two points $\vr_1$ and $\vr_2$ need
not be on $\beta$ for this contribution to $\delta G_{\rm AA}$. 
If both $\vr_1$ and $\vr_2$ are not on
$\beta$, as in Fig.\ \ref{fig:2}b, 
the trajectories $\beta$ and $\alpha_2$ undergo a
``small angle encounter''. 
Small angle encounters play a crucial role in the theory of weak 
localization and the shot noise power in ballistic conductors \cite{kn:aleiner1996,kn:richter2002,kn:agam2000,kn:braun2006,kn:whitney2006,kn:brouwer2007b}.
The condition that the action difference
$\Delta {\cal S} = S_{\alpha_1} - {\cal S}_{\alpha_2} + {\cal 
S}_{\alpha_3} - {\cal S}_{\beta}$ be of order $\hbar$ translates 
to the condition that the duration of the encounter be equal to the
Ehrenfest time $\tau_{\rm E}$ \cite{kn:aleiner1996,kn:richter2002}.
The same is true if one of the points
$\vr_1$ and $\vr_2$ is on $\beta$ and one is not, as in Fig.\
\ref{fig:2}c and d. If both points $\vr_1$ and $\vr_2$ are on $\beta$,
as in Fig.\ \ref{fig:2}e,
the travel time between $\vr_1$ and $\vr_2$ must be less than
$\tau_{\rm E}$ \cite{kn:whitney2006}.


%

The configurations of Figs.\ \ref{fig:2}a and e are essentially one
dimensional. Their contributions are found to cancel precisely if the
travel time $\tau_{\alpha_2}$ between $\vr_1$ and $\vr_2$ is less than
$\tau_{\rm E}$. 
The configurations of Figs.\ \ref{fig:2}b--d require a
summation over trajectories involved in a small angle encounter,
taking into account the proper action difference $\Delta {\cal S}$. 
For the present case, where the interaction propagator
${\cal D}(\vr_1,\vr_2;\omega)$ is independent of the precise location
of the coordinates $\vr_1$ and $\vr_2$ inside each quantum dot, this
summation is essentially identical to that needed for the calculation 
of the shot noise power. Following the method
of Refs.\ \cite{kn:whitney2006,kn:braun2006,kn:brouwer2007b}, 
one then finds
\begin{widetext}
\begin{eqnarray}
  \delta G_{\rm AA} &=& \frac{e^2}{h}
  \int \frac{d \omega}{i \hbar}
  {\cal F}(\omega/T)
  \sum_{m, n} \tilde {\cal D}^{\rm A}_{mn}(\omega) \nu_n
  \left\{
  \int_{\tau_{\rm E}}^{\infty} dt\, P^{\rm L}_m
  P_{mn}^{\vphantom{remove this}}(t) P^{\rm R}_{n} e^{-i \omega t/\hbar}
  \right. \nonumber \\ && \ \ \ \ \mbox{}  -
   \sum_{k,l} \left. \int_0^{\infty} dt_2 dt_1 
  [\partial_{\tau_{\rm E}} + \delta(t_1) + \delta(t_2)]
  P_{mk}^{\vphantom{remove this}}(t_2) P^{\rm L}_k
  P_{kl}^{\vphantom{remove this}}(\tau_{\rm E}) 
  P^{\rm R}_l P_{ln}^{\vphantom{remove this}}(t_1) e^{-i \omega (\tau_{\rm E}+t_1+t_2)/\hbar}
  \right\}
  + \mbox{c.c.},
  \label{eq:dGAA1}
\end{eqnarray}
where ${\cal F}(x) = (\sinh x - x)/(\cosh x - 1)$,
$P^{\rm L}_{n}$ and $P^{\rm R}_{n} = 1 - P^{\rm L}_n$ are
the classical probabilities that an electron in dot $n$ exits the
system through the left or right contacts,
respectively, and $P_{mn}(t)$ is the classical probability that 
an electron in dot $n$ is in dot $m$ after a time
$t$. The first term in Eq.\ (\ref{eq:dGAA1}) comes from trajectory
configurations of Fig.\ \ref{fig:2}a and e; the second term comes from
Fig.\ \ref{fig:2}b--d. In matrix language, one has $P_{kl}(t) = 
[\exp(-\tilde G t /e^2 \tilde \nu)]_{kl}$, 
where the $2 \times 2$ matrices $\tilde G$ and $\tilde \nu$ were 
defined below 
Eq.\ (\ref{eq:Dprop}).

In the special case of a symmetric double quantum dot with
$G_1 = G_2 \equiv G_{\rm d}$
and $\nu_1 = \nu_2 \equiv \nu$, and neglecting the first term in
Eq.\ (\ref{eq:Dprop}), Eq.\ (\ref{eq:dGAA1}) 
simplifies to
\begin{eqnarray}
  \delta G_{\rm AA} &=&
  \frac{e^2}{\pi \hbar} 
  \frac{G_{\rm d} G_{\rm c}^2
  (\tau_{{\rm D}-} e^{-\tau_{\rm E}/\tau_{{\rm D}+}} +
  \tau_{{\rm D}+} e^{-\tau_{\rm E}/\tau_{{\rm D}-}})}
  {(G_{\rm d} + 2 G_{\rm c})^3}\,
  \mbox{Im}\,
  \int d\omega
  \frac{e^{-i \omega \tau_{\rm E}} {\cal F}(\hbar\omega/T)}
  {(1 + i \omega \tau_{{\rm D}+})(1 + i \omega \tau_{{\rm D}-})},
  \label{eq:dGomega}
\end{eqnarray}  
\end{widetext}
where $\tau_{\rm D+} = e^2 \nu/G_{\rm d}$ 
and $\tau_{\rm D-} = e^2 \nu/(G_{\rm d} + 2 G_{\rm c})$ 
are the characteristic
dwell times of the double quantum dot.
The indices $+$ ($-$) refer to relaxation of (anti)symmetric
charge configurations. 
The frequency integral in Eq.\ (\ref{eq:dGomega})
can not be performed in closed form, except in asymptotic limits. If
$\tau_{\rm E} \ll \min(\tau_{{\rm D}\pm},\hbar/T)$, one
recovers the result of random matrix theory \cite{kn:kupferschmidt2007},
\begin{eqnarray}
  \delta G_{\rm AA} &=&
  - \frac{2 e^2}{\pi \hbar} 
  \frac{G_{\rm d} G_{\rm c}^2}{(G_{\rm d} + 2
  G_{\rm c})^3}
  \frac{\tau_{{\rm D}+} + \tau_{{\rm D}-}}
       {\tau_{{\rm D}+} - \tau_{{\rm D}-}}
  \ln \frac{\tau_{{\rm D}+}}{\tau_{{\rm D}-}}
\end{eqnarray}
if $\tau_{\rm D} T \ll \hbar$, and
\begin{eqnarray}
  \delta G_{\rm AA} &=&
  - \frac{e^2}{3 T} 
  \frac{G_{\rm d} G_{\rm c}^2}{(G_{\rm d} + 2
  G_{\rm c})^3}
  \frac{\tau_{{\rm D}+} + \tau_{{\rm D}-}}{\tau_{{\rm D}+} \tau_{{\rm D}-}}
\end{eqnarray}
if $\tau_{\rm D} T \gg \hbar$.
%
As soon as $\tau_{\rm E}$ becomes comparable to $\tau_{\rm D}$ or
$\hbar/2 \pi T$, $\delta G_{\rm AA}$ acquires a
dependence on $\tau_{\rm E}$, which becomes exponential
in the limit of large Ehrenfest times,
\begin{equation}
  \delta G_{\rm AA} = - \frac{2 e^2}{\pi \hbar \tau_{\rm E}}
    \frac{G_{\rm d} G_{\rm c}^2
  (\tau_{{\rm D}-} e^{-\tau_{\rm E}/\tau_{{\rm D}+}} +
  \tau_{{\rm D}+} e^{-\tau_{\rm E}/\tau_{{\rm D}-}})}
  {(G_{\rm d} + 2 G_{\rm c})^3}
  \label{eq:dGAA4}
\end{equation}
if $\tau_{\rm D} \ll \tau_{\rm E} \ll \hbar/T$, and
\begin{eqnarray}
  \label{eq:dGAA5}
  \delta G_{\rm AA} &=&
  - \frac{2 e^2}{\pi \hbar}
  \frac{G_{\rm d} G_{\rm c}^2\tau_{\rm E} e^{-2 \pi
  T \tau_{\rm E}/\hbar}}
  {(G_{\rm d} + 2 G_{\rm c})^3}  
  \\ && \mbox{} \times
  \frac{
  \tau_{{\rm D}-} e^{-\tau_{\rm E}/\tau_{{\rm D}+}} +
  \tau_{{\rm D}+} e^{-\tau_{\rm E}/\tau_{{\rm D}-}}}
  {(\tau_{{\rm D}+} + \hbar/2 \pi T)
  (\tau_{{\rm D}-} + \hbar/2 \pi T)}.
  \nonumber
\end{eqnarray}
if $\tau_{\rm E} \gg \hbar/T$. Equations (\ref{eq:dGAA4}) and
(\ref{eq:dGAA5}) reproduce the general exponential dependence of Eq.\
(\ref{eq:exponential}), with $\tau_{\rm D} = \tau_{\rm D+}$ taken to
be the larger of the two characteristic dwell times. For a generic
(non-symmetric) double
quantum dot one finds the same dependence on $\tau_{\rm
E}$ from the general expression of Eq.\ (\ref{eq:dGAA1}).
  
Let us now compare the interaction correction $\delta G_{\rm AA}$ to
the weak localization correction $\delta G_{\rm WL}$ in the same
system, for which one finds \cite{kn:brouwer2007b}
\begin{eqnarray}
  \delta G_{\rm WL} &=&
  \frac{2 e^2}{h} \sum_{k,l}
  \partial_{\tau_{\rm E}}
  P^{\rm L}_{k} P^{\rm R}_{k}
  P_{kl}^{\vphantom{omit this}}(\tau_{\rm E})
  \int dt P_{ll}^{\vphantom{omit this}}(t),
\end{eqnarray}
in the absence of a magnetic field. (With a magnetic field, $\delta
G_{\rm WL} = 0$.)
For the special case of a symmetric double quantum dot this simplifies to
\begin{eqnarray}
  \delta G_{\rm WL} &=&
  - \frac{2 G_{\rm c} (G_0 + G_{\rm c})^2 e^{-\tau_{\rm E}/\tau_{\rm
  D+}}}
  {(G_0 + 2 G_{\rm c})^3}.
  \label{eq:WLspecial}
\end{eqnarray}
Comparing with the expressions for $\delta G_{\rm AA}$ derived above,
we note that both $\delta G_{\rm WL}$ and $\delta
G_{\rm AA}$ disappear $\propto \exp(-\tau_{\rm E}/\tau_{\rm D})$
in the limit $\tau_{\rm E} \gg
\tau_{\rm D}$ at zero temperature. 
[The fact that $\delta G_{\rm WL}$ has a single
exponential decay rate in Eq.\ (\ref{eq:WLspecial})
while $\delta G_{\rm AA}$ has two exponential
decay rates is an artifact of the symmetry $G_1 = G_2$.]
At finite temperature, $\delta G_{\rm WL}$ and $\delta G_{\rm AA}$
have a different dependence on $\tau_{\rm E}$, since 
$\delta G_{\rm WL}$ has no explicit 
temperature dependence through the thermal time $\hbar/2 \pi
T$.


The exponential dependence of $\delta G_{\rm AA}$ on $\tau_{\rm
  E}/\tau_{\rm D}$ and $\tau_{\rm E} T/\hbar$ [Eq.\
(\ref{eq:exponential})] is the main result of this letter. Although
it was derived by explicit calculation of $\delta G_{\rm AA}$ in a
ballistic double quantum dot, we should point out that the {\em
origin} of the $\tau_{\rm E}$ dependence of $\delta G_{\rm AA}$ is 
in the structure of the classical trajectories contributing to 
$\delta G_{\rm AA}$ shown in Fig.\ \ref{fig:2}, not in the details of
the semiclassical calculation.
It is because of this that we expect that 
our qualitative conclusions for the $\tau_{\rm E}$ dependence of 
$\delta G_{\rm AA}$
carry over to other geometries, notwithstanding
differences in the quantitative evaluation of $\delta G_{\rm AA}$ in
those cases \cite{foot2}.

We thank I.~Aleiner for useful discussions. This work was supported by
the Cornell Center for Materials research under NSF grant no.\ DMR 0520404,
the Packard Foundation, the Humboldt Foundation, and by the NSF under
grant nos.\ DMR 0334499 and 0705476.


\begin{thebibliography}{26}
\bibitem{foot2}
This difference is more than the mere adaptation 
of Eq.\
(\ref{eq:dGAA1}) to a different classical dynamics, as one 
can no longer rely on the effectively long-range form of 
the electron-electron interaction in a general setting. This
property of the interaction entered our calculation in two
ways: It suppressed the Hartree 
contribution to $\delta G_{\rm AA}$ and simplified the trajectory
summations leading up to Eq.\
(\ref{eq:dGAA1}). A theory of the Altshuler-Aronov correction in an
arbitrary ballistic conductor would have to deal with both effects.
\end{thebibliography}

\end{document}